\documentclass{emulateapj}
\usepackage{graphicx}
\usepackage{natbib}

\slugcomment{Accepted by ApJL, draft from \today}

\shorttitle{Magnetic field structure in a high-mass outflow/disk system}
\shortauthors{Beuther et al.}

\begin{document}


\title{Magnetic field structure in a high-mass outflow/disk system}


\author{H.~Beuther\altaffilmark{1}, W.H.T.~Vlemmings\altaffilmark{2}, R.~Rao\altaffilmark{3}, F.F.S.~van der Tak\altaffilmark{4}}





\altaffiltext{1}{Max-Planck-Institute for Astronomy, K\"onigstuhl 17,
              69117 Heidelberg, Germany, beuther@mpia.de}
\altaffiltext{2}{Argelander Institute for Astronomy, University of Bonn, 
              Auf dem Hügel 71, 53121 Bonn, Germany, wouter@astro.uni-bonn.de}
\altaffiltext{3}{Academia Sinica, Institute of Astronomy and Astrophysics, 
             Taipei, Taiwan {\it and} Submillimeter Array, 645 N. 
             Aohoku Pl, Hilo, HI 96720, USA, rrao@sma.hawaii.edu}
\altaffiltext{4}{SRON Netherlands Institute for Space Research, 
             Landleven 12, 9747 AD Groningen, The Netherlands, vdtak@sron.nl}
           

\begin{abstract}
  To characterize the magnetic field structure of the outflow and core
  region within a prototypical high-mass star-forming region, we
  analyzed polarized CO(3--2) -- for the first time observed with the
  Submillimeter Array -- as well as 880\,$\mu$m submm continuum
  emission from the high-mass outflow/disk system IRAS\,18089-1732.
  Both emission features with polarization degrees at a few percent
  level indicate that the magnetic field structure is largely aligned
  with the outflow/jet orientation from the small core scales to the
  larger outflow scales. Although quantitative estimates are crude,
  the analysis indicates that turbulent energy dominates over magnetic
  energy. The data also suggest a magnetic field strength increase
  from the lower-density envelope to the higher-density core.
\end{abstract}



\section{Introduction}
\label{intro}

The influence of magnetic fields on star formation processes has been
intensely discussed since decades (e.g., \citealt{crutcher2005}). For
magnetically supported molecular clouds, the magnetic field structure
is expected to deform during the cloud collapse, and theory predicts
an hour-glass shaped magnetic field structure pinched toward the
protostar (e.g., \citealt{stahler2005}).  This pinched magnetic field
morphology is also required in the classical jet-launching scenario,
where the gas is ejected from the accretion disk along the bent
magnetic field lines (e.g., \citealt{blandford1982,camenzind1990}).
Submillimeter Array (SMA) observations of linearly polarized dust
continuum emission toward the low-mass protostar NGC1333 IRAS4 have
clearly shown the pinched hour-glass shaped magnetic field field
morphology \citep{girart2006}.  The quest for magnetic fields is as
important in high-mass star formation.  Recent submm dust polarization
observations have revealed similarly shaped magnetic field
morphologies toward two massive star-forming regions
\citep{girart2009,tang2009a}.  In addition to that, molecular outflows
are known to be as ubiquitous in high- as in low-mass star formation,
and if the launching mechanisms are of similar nature, magnetic fields
have to be present in massive cores as well.

While most previous high-spatial-resolution magnetic field studies
were either based on dust continuum polarization observations or maser
studies (e.g., \citealt{girart2009,vlemmings2008}), the molecular gas
can also show non-maser spectral line polarization signatures due to
the Goldreich-Kylafis effect \citep{goldreich1981,goldreich1982}.
Observations of this effect are particularly promising for studying
the magnetic field in molecular outflows
\citep{girart1999,lai2003,cortes2005,cortes2008}.

We present a combined polarization study of the 880\,$\mu$m submm
continuum and the CO(3--2) emission toward the well-studied massive
disk-outflow system IRAS\,18089-1732. The luminosity and gas mass of
the region at a distance of 3.6\,kpc are $10^{4.5}$\,L$_{\odot}$ and
1200\,M$_{\odot}$, respectively
\citep{sridha,beuther2002a,williams2004}.
\citet{vlemmings2008b} derives for this region a line-of-sight
magnetic field strength of 8.4\,mG from Zeeman splitting measurements
of the 6.7\,GHz CH$_3$OH maser line at densities $>10^6$\,cm$^{-3}$.
SMA studies revealed a molecular outflow in approximately north-south 
direction and rotational signatures in the dense gas perpendicular to
the outflow which were confirmed by high-excitation NH$_3$
observations \citep{beuther2005c,beuther2008a}. Since IRAS\,18089-1732
remains a single compact source up to the highest angular resolution
($<1''$), and since the outflow-disk orientation is well constrained,
this region is the ideal target for magnetic field investigation in
high-mass star formation.

\section{Observations} 
\label{obs}

The region was observed with the SMA\footnote{The Submillimeter Array
  is a joint project between the Smithsonian Astrophysical Observatory
  and the Academia Sinica Institute of Astronomy and Astrophysics, and
  is funded by the Smithsonian Institution and the Academia Sinica.})
in the compact and extended configurations in May 2008 and August
2009. The phase center was R.A.~18:11:51.4 and Dec.: -17:31:28.5
(J2000.0) and the tuning frequency 345.796\,GHz in the upper sideband
($v_{\rm{lsr}}=33.8$\,km\,s$^{-1}$).  While the compact array
observations were still conducted in the old setup with 2\,GHz
bandwidth in both sidebands, the extended configuration data profited
from the increased bandwidth of 4\,GHz in both sidebands. The spectral
resolution in both cases was 0.8125\,kHz ($\sim 0.7$\,km/s). The
weather was good with zenith opacities mostly below 0.25 at 346\,GHz.
Calibration and imaging were conducted in MIRIAD \citep{sault1995}.
The passband calibration was derived from 3c454.3, and fluxes were
calibrated via the regularly monitored SMA calibrator database. The
flux density scale is estimated to be accurate within 15\%. For the
phase and amplitude calibration regularly interleaved observations of
the quasar 1733-130 were conducted. The polarimeter system of the SMA
is described in detail in \citet{marrone2006,marrone2008}. As
polarization calibrator we used for the compact and extended
configuration 3c273 and 3c454.3, respectively. For the CO(3-2)
polarization calibration we took only the emission from the 7 chunks
of the bandpass around the CO(3--2) line (approximately 700\,MHz). For
the continuum data we used both configurations resulting in a
synthesized beam of $1.65''\times 1.05''\,(\rm{P.A.}\,51^{\circ})$.
Since the CO(3--2) emission is more extended, we used only data from
the compact configuration resulting in a synthesized beam of
$2.11''\times 1.34''\,(\rm{P.A.}\,52^{\circ})$. The 1$\sigma$ rms
values of the continuum Stokes I and Stokes Q/U images are
18.4\,mJy\,beam$^{-1}$ and 4.3\,mJy\,beam$^{-1}$, respectively,
because the noise is dominated by side-lobes from strong sources due
to insufficient uv-sampling. For the CO(3--2) emission the 1$\sigma$
values at 4\,km\,s$^{-1}$ spectral resolution of the Stokes I and
Stokes Q/U images are 140\,mJy\,beam$^{-1}$ and 75\,mJy\,beam$^{-1}$,
respectively. The polarized continuum and CO(3--2) images were
produced applying a 3$\sigma$ cutoff to the data. For comparison, for
the continuum we also show the results applying 2$\sigma$ cutoffs in
Fig.~\ref{continuum}.  From the other spectral lines in the setup,
here we only present the integrated SiO(8--7) emission
(Fig.~\ref{continuum}) to outline the direction of the outflow/jet.
The synthesized beam and the rms of the SiO(8--7) image integrated
from 30 to 40\,km\,s$^{-1}$ are $2.11''\times
1.33''\,(\rm{P.A.}\,52^{\circ})$ and 52\,mJy\,beam$^{-1}$,
respectively.

\section{Results}

\subsection{Submm dust continuum polarization}

\begin{figure}
\includegraphics[angle=-90,width=9cm]{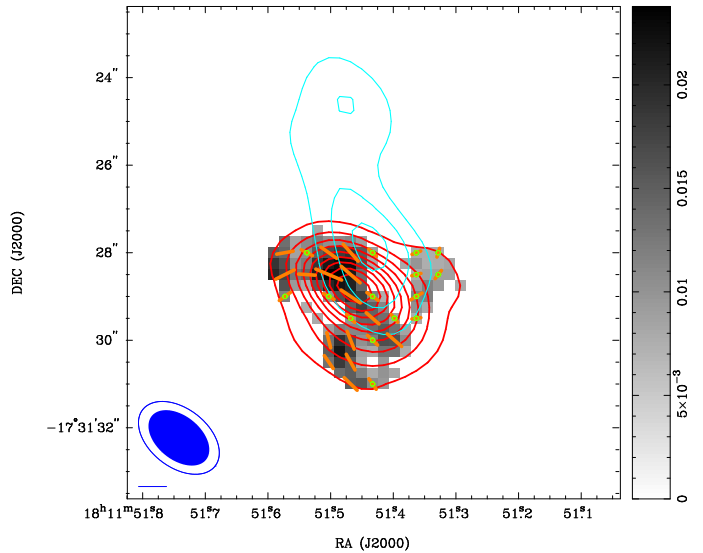}
\caption{The grey-scale presents the linearly polarized 880\,$\mu$m
  continuum image of IRAS\,18089-1732 in units of Jy\,beam$^{-1}$.
  The line-segments plotted in approximate half-beam spacing of
  0.5$''$ show the magnetic field orientation assuming that it is
  oriented perpendicular to the polarization (line-segment length
  scales with strength of polarized emission). Line-segments marked by
  a small green circle are those where a 2$\sigma$ cutoff was applied.
  The red contours present the integrated Stokes I image of the
  continuum emission in 10$\sigma$ steps. To outline the outflow
  direction, the blue contours show the SiO(8--7) emission integrated
  from 30 to 40\,km\,s$^{-1}$ ($5\sigma$ steps of
  260\,mJy\,beam$^{-1}$). Full and open ellipses present the
  synthesized beams of the continuum and SiO emission.}
\label{continuum}
\end{figure}

Figure \ref{continuum} presents the linearly polarized
($\sqrt{Q^2+U^2}$) as well as Stokes I 880\,$\mu$m continuum emission
toward the outflow/disk system IRAS\,18089-1732.  While the Stokes I
image exhibits the same single-peaked structure as known from previous
observations (e.g., \citealt{beuther2005c}), the linearly polarized
emission has sub-structure offset from the main peak. The strongest
linearly polarized emission with a peak flux of 23.9\,mJy\,beam$^{-1}$
is $\sim 0.7''$ N-NE (north-northeast) of the Stokes I peak, and the
polarized emission shows additional substructure south of the Stokes I
peak position. At the linearly polarized peak position
(R.A.\,(J2000.0) 18:11:51.51, Dec.\,(J2000.0) -17:31:28.5), the polarization
fraction is at a $\sim$1.5\% level, at the Stokes I peak position
(R.A.\,(J2000.0) 18:11:51.47, Dec.\,(J2000.0) -17:31:28.8), the polarization
fraction is still at a $\sim$1.1\% level (Table \ref{table}). The
Stokes I peak and integrated fluxes within the 10$\sigma$ contours are
1.93\,mJy\,beam$^{-1}$ and 4.24\,Jy, at 880\,$\mu$m wavelength,
respectively. Following \citet{hildebrand1983}, assuming optically
thin dust continuum emission with a dust emissivity index $\beta =2$
at a temperature of $\sim$100\,K (e.g., \citealt{beuther2004b}) and a
gas-to-dust ratio of 186 (e.g., \citealt{draine2007}), the derived gas
column densities $N$ and masses $M$ are $9.5\times 10^{24}$\,cm$^{-2}$
and 197\,M$_{\odot}$, respectively.  Taking into account that in
particular the mass is a lower limit due to the interferometric
missing flux, clearly we are dealing with a high column density and
massive core. Assuming a spherical distribution with an approximate
core diameter of $3.5''$ (linear size of $\sim 12600$\,AU), we get an
approximate average core density $n$ of $5\times 10^7$\,cm$^{-3}$.

For elliptical dust grains, the polarization orientation of the
radiation field is expected to be perpendicular to the magnetic field
if the dust grains spin fast enough.  Therefore, to infer information
about the magnetic field structure, we rotated the line-segment
orientation of the polarized emission in Figure \ref{continuum} by
90$^{\circ}$ (line-segments are shown in approximate Nyquist sampling
of $0.5''$, indicating the number of independent measurements). In
this picture, the overall orientation of the magnetic field component
in the plane of the sky follows approximately the general
morphological structure of the polarized emission with an orientation
roughly along an axis in N-NE to S-SW direction. This general
orientation is approximately aligned with the outflow traced by the
SiO(8--7) emission (Fig.~\ref{continuum} and \citealt{beuther2004b}).
  Toward the western edge of the core, the field orientation bends a
  little bit toward the mid-plane of the underlying accretion disk
  \citep{beuther2008a}.

\subsection{Polarization of the CO(3--2) 
outflow emission}

\begin{figure}
\includegraphics[angle=-90,width=9cm]{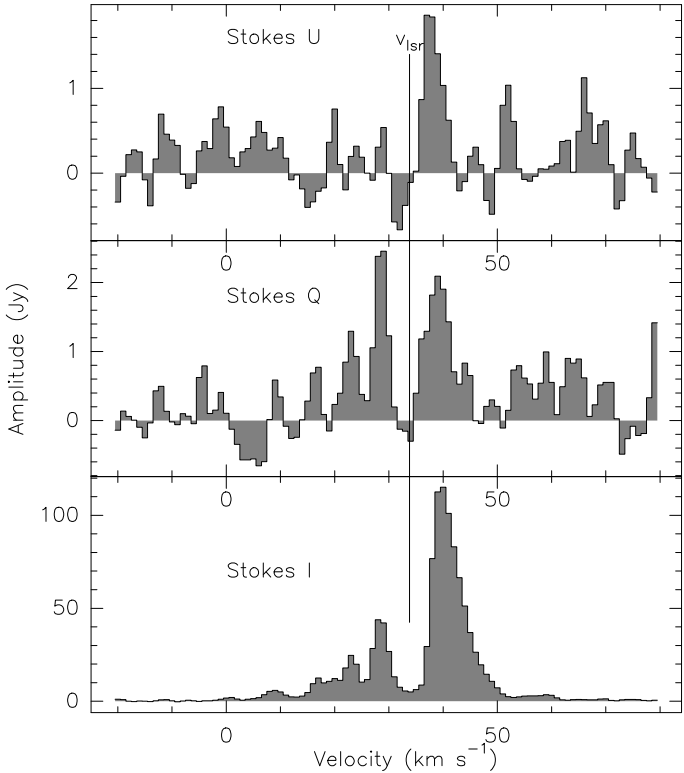}
\caption{The three panels present vector-averaged spectra of the
  CO(3--2) line taken on the shortest baseline of the Stokes I, Q and
  U components. The vertical line marks the $v_{\rm{lsr}}$.}
\label{spectra} \end{figure}

\begin{figure*}
  \includegraphics[angle=-90,width=18cm]{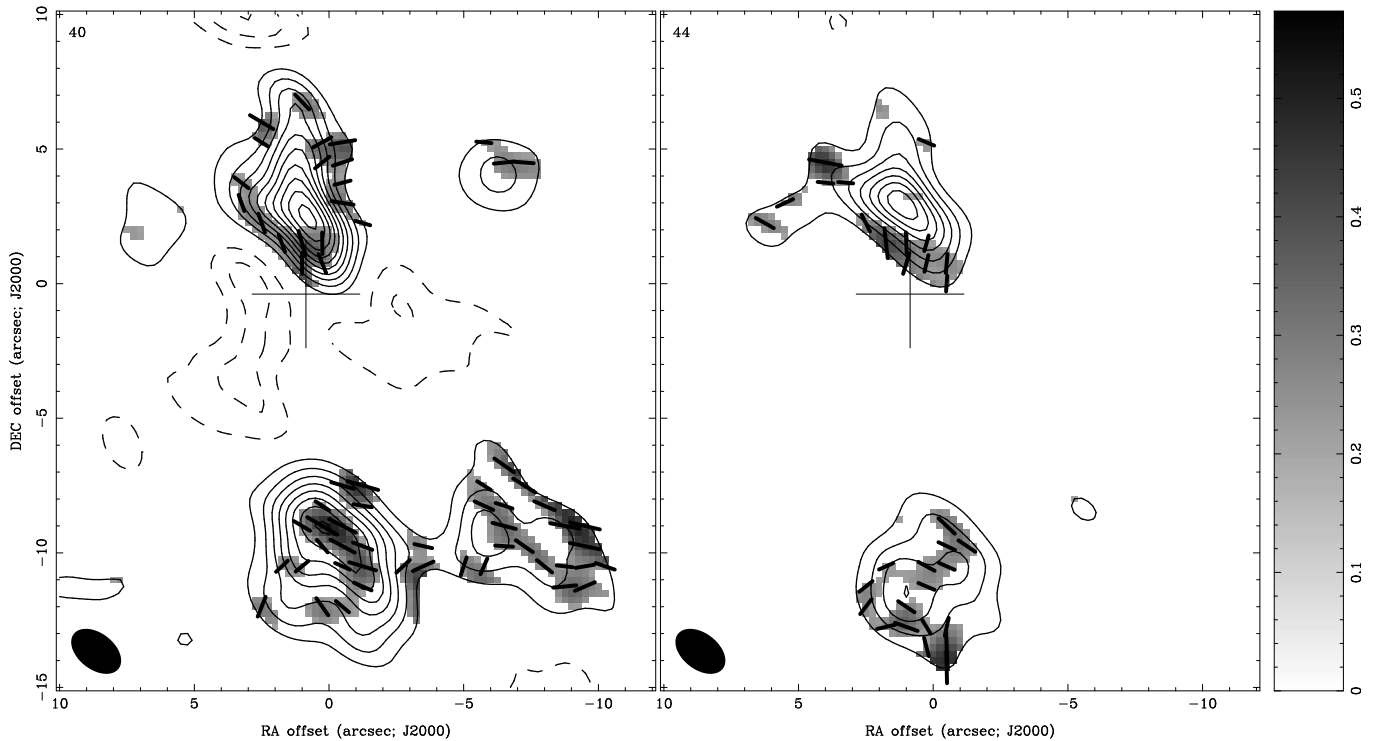} \caption{The
    grey-scale presents the linearly polarized CO(3--2) emission in
    two 4\,km\,s${-1}$ wide channels in units of Jy\,beam$^{-1}$
    (central velocities are marked in the top-left corners). The
    line-segments plotted in approximate half-beam spacing of 0.75$''$
    show the orientation of the polarized emission (line-segment
    length scales with strength).  The contours present the
    corresponding CO(3--2) Stokes I image in 10$\sigma$ steps
    (positive and negative features as full and dashed contours). The
    cross marks the position of the submm continuum peak.  The
    synthesized beam is shown at the bottom left of each panel.}
  \label{co} \end{figure*}

Furthermore, we observed simultaneously the polarized emission of
CO(3-2). Figure \ref{spectra} presents the Stokes I and the linearly
polarized Stokes U and Q spectra obtained vector-averaged on the
shortest baseline. While the emission at the $v_{\rm{lsr}}$ is
filtered out in all cases, the velocity component around
40\,km\,s$^{-1}$ is not only detected in Stokes I but also in both
linearly polarized components of Stokes U and Q.
From these data, we derive spatially and spectrally resolved images of
the total linearly polarized CO(3--2) emission in this region
($\sqrt{Q^2+U^2}$). Figure~\ref{co} shows two 4\,km\,s$^{-1}$ wide
channels of polarized CO(3--2) emission around the velocity of
40\,km\,s$^{-1}$. We again clearly identify the overall N-NE to S-SW
outflow direction, but here not only the northern component but also a
southern counterpart.  Since we do not identify a clear
blue-red-shifted dichotomy, the outflow is likely close to the plane
of the sky. The orientation of the polarization follows mainly the
outflow direction, and we find polarization degrees in the images of
the order of 10\% (Table \ref{table}), higher than found toward other
regions (e.g., \citealt{cortes2005}).

\subsection{Magnetic field strength}

Although magnetic field estimates based on polarization data are
relatively unreliable since they are based on a statistical analysis
of the polarization line-segments, nevertheless, we have enough
independent polarization measurements in the continuum and line
emission that at least rough estimates are useful.
\paragraph{Dust continuum polarization} Following the work by
\citet{chandrasekhar1953} and more recently \citet{girart2006}, we
estimate the strength of the magnetic field component in the plane of
the sky $B_{\rm{pos}}\approx Q\sqrt{4\pi \rho}\frac{\Delta
  v_{\rm{los}}}{\Delta \phi}$ with $\rho$ the average density derived
above in g\,cm$^{-3}$, $\Delta v_{\rm{los}}$ the approximate 1D
velocity dispersion along the line of sight ($\Delta v_{\rm{los}}\sim
\Delta v(\rm{CH_3CN})/\sqrt{8ln2}\sim 2.2$\,km\,s$^{-1}$ with $\Delta
v(\rm{CH_3CN})\sim 5.1$\,km\,s$^{-1}$, \citealt{beuther2005c}) and
$\Delta \phi$ the approximate flux-weighted dispersion of the
polarization angle line-segments of $\sim$27$^{\circ}$. In practice,
we use the rms of flux-weighted line-segments ignoring any bending of
the field lines which implies that the inferred B-values should be
lower limits.  $Q$ is a dimensionless parameter accounting for the
cloud structure, and work by \citet{ostriker2001} suggests $Q\sim 0.5$
for turbulent clouds.  Since the density $\rho$ depends on the
distance $d$, and the latter is the relatively uncertain kinematic
distance, one should keep in mind an additional $B$-dependence of
$B\propto 1/\sqrt{d/3.6\rm{kpc}}$. The derived magnetic field strength
in the plane of the sky is $B_{\rm{pos}}\approx 11$\,mG. Because
\citet{vlemmings2008b} derives a comparable line-of-sight magnetic
field strength for similar densities of the gas from maser Zeeman
observations ($B_{\rm{los}}\approx 8$\,mG), the total magnetic field
strength at the given core density should be $B_{\rm{tot}}\approx
\sqrt{B_{\rm{pos}}^2+B_{\rm{los}}^2}\approx 14$\,mG.

While this is larger than the value previously derived in DR21 (1\,mG,
albeit measured at a density more than an order of magnitude lower,
\citealt{lai2003}), it is of the same order as the recent estimates
based on dust polarization measurements for G31.41 \citep{girart2009}
and those derived from CH$_3$OH maser Zeeman measurements towards a
sample of high-mass star-forming regions by
\citet{vlemmings2008b,vlemmings2010,surcis2009}. The corresponding
Alfvenic velocity is $v_A\propto B/\sqrt{\rho} \approx
3.1$\,km\,s$^{-1}$. The important mass-to-flux ratio can be estimated
via $M/\Phi_B\approx 7.6\times 10^{-24} \frac{N(\rm{H_2})}{B}\approx
5.2$ in units of the critical mass-to-flux ratio
$(M/\Phi_B)_{\rm{crit}}$ ($N(\rm{H_2})$ and $B$ are given in cm$^{-1}$
and mG, \citealt{crutcher1999,troland2008}). While the latter estimate
indicates that the magnetic field cannot prevent the core from
collapse, comparison of the Alfvenic velocity with the 1D velocity
dispersion suggests that the collapse proceeds at approximately or
slightly below the Alfvenic speed. Furthermore following
\citet{girart2009}, we estimate the ratio of turbulent to magnetic
energy via $\beta\approx 3(\Delta v_{\rm{los}}/v_{A_{los}})^2\approx
2.5$ (with $v_{A_{los}}\approx 2.4$\,km\,s$^{-1}$ the Alfvenic
velocity along the line of sight). Because the estimated
$B_{\rm{pos}}$ is a lower limit, the ratios $M/\Phi_B$ and $\beta$ are
upper limits.  While for G31.4, W75N, W51 and CepA the magnetic field
may play a dominant role determining the dynamics of the systems
\citep{girart2009,surcis2009,tang2009a,vlemmings2010}, for
IRAS\,18089-1732 the system appears to be more strongly dominated by
turbulent motions.

\paragraph{CO(3--2) polarization}

Following the same Chandrasekhar-Fermi approach as above, we derive a
very rough estimate of the magnetic field strength within the outflow
lobes. The rms of the polarization angle distribution in the northern
lobe is $\Delta \phi \sim 45^{\circ}$ (again measured as the rms of
flux-weighted line-segments ignoring any bending of the field lines).
However, getting a density estimate based on the given data is far
more difficult. Since we cannot derive the density ourselves, we can
only assume an average density in the more diffuse envelope where the
outflow is observed.  Average densities of order $10^3$\,cm$^{-3}$
should approximately represent the envelope/outflow.  With these
values and the same line width as for the core, we get magnetic field
estimates of order $\sim 28$\,$\mu$G, about three orders of magnitude
below the central core magnetic field values.  While the
Chandrasekhar-Fermi method in principle is prone to many
uncertainties~-- e.g., the $\Delta \phi$ requires subtraction of the
ordered magnetic field~-- it may even be less applicable in the case
of the outflow lobes since the outflow may determine most of the
polarization orientation. Furthermore, the density uncertainties add
to the total uncertainty of the magnetic field estimate.  While we
cannot give a reliable estimate of this total uncertainty, it is
nevertheless likely below 3 orders of magnitude. Hence, the magnetic
field strength in the outflow region appears lower than in the core
region. This is consistent with previous results for DR21, where the
CO polarization traces magnetic fields of the order 10\,$\mu$G in the
envelope and the dust polarization field strengths of the order mG in
the denser core regions \citep{cortes2005,lai2003}.

\section{Discussion and conclusion}
\label{general}

The combined polarization data of the submm continuum and the CO(3--2)
emission allow us to study the magnetic field structure in the core
and outflow of the high-mass disk-outflow system IRAS\,18089-1732.
For the dust continuum emission, compared to magnetic field hour glass
shape structures like those found in NGC1333 or G31.41
\citep{girart2006,girart2009}, the case is less clear here.  Inferring
an hour-glass shape like structure indicative of ambipolar diffusion
from these data would likely be an over-interpretation.  However, in a
broader sense this may not be such a surprise. Already
\citet{mouschovias1979} pointed out that adding rotation to the
collapsing core could destroy the hour-glass shape structure.
Therefore, in a statistical sense it may be that finding the
hour-glass signatures could be more the exception than the rule. An
analysis of a larger sample of sources is required to confirm or
reject this hypothesis.  Nevertheless, the data of IRAS\,18089-1732
are consistent with a picture where the outflow and magnetic field
orientation are tightly linked.

For the CO(3--2) emission, since the Goldreich-Kylafis effect occurs
only when the line emission is dominated by radiative excitation and
not collisions, it sensitively depends on the anisotropy of the
optical depth and the radiation field as well as on the orientation of
the magnetic field (for a qualitative descriptions see
\citealt{kylafis1983}).  Additionally, there exists an ambiguity
whether the polarization orientation is perpendicular or parallel to
the magnetic field.  However, \citet{cortes2005} have shown that a
large radiation and optical depth anisotropy produces high degrees of
polarization parallel to the magnetic field (their Fig.~7). Since we
observe such a high polarization degree of order 10\%, and since the
radiation and velocity field is dominated by an anisotropic outflow
cone with a central protostellar radiation source, this scenario
supports an interpretation that the magnetic field and CO(3--2)
polarization orientation in IRAS\,18089-1732 are spatially aligned.
This is consistent with previous CO polarization studies in DR21 and
NGC1333 \citep{girart1999,lai2003} as well as with the picture of
magnetic field accelerated and collimated jets where the magnetic
field and outflow should have the same orientation.

To the authors' knowledge, this is the first successful detection of
the Goldreich-Kylafis effect in the CO(3--2) line. Comparing the
outflow morphology with the polarization direction, it is likely that
the magnetic field and polarization both are aligned with the outflow.
Although the Chandrasekhar-Fermi method allows only rough magnetic
field strength estimates, the data are indicative of a magnetic field
strength increase from the envelope toward the core. Furthermore, the
observations indicate that turbulent energy likely dominates over
magnetic energy.

\begin{acknowledgements} 
W.V. acknowledges support by the Deutsche Forschungsgemeinschaft through
the Emmy Noether Research grant VL 61/3-1.
\end{acknowledgements}


\begin{thebibliography}{34}
\expandafter\ifx\csname natexlab\endcsname\relax\def\natexlab#1{#1}\fi

\bibitem[{{Beuther} {et~al.}(2004){Beuther}, {Hunter}, {Zhang}, {Sridharan},
  {Zhao}, {Sollins}, {Ho}, {Ohashi}, {Su}, {Lim}, \& {Liu}}]{beuther2004b}
{Beuther}, H., {Hunter}, T.~R., {Zhang}, Q., {et~al.} 2004, \apjl, 616, L23

\bibitem[{{Beuther} {et~al.}(2002){Beuther}, {Schilke}, {Menten}, {Motte},
  {Sridharan}, \& {Wyrowski}}]{beuther2002a}
{Beuther}, H., {Schilke}, P., {Menten}, K.~M., {et~al.} 2002, \apj, 566, 945

\bibitem[{{Beuther} \& {Walsh}(2008)}]{beuther2008a}
{Beuther}, H. \& {Walsh}, A.~J. 2008, \apjl, 673, L55

\bibitem[{{Beuther} {et~al.}(2005){Beuther}, {Zhang}, {Sridharan}, \&
  {Chen}}]{beuther2005c}
{Beuther}, H., {Zhang}, Q., {Sridharan}, T.~K., \& {Chen}, Y. 2005, \apj, 628,
  800

\bibitem[{{Blandford} \& {Payne}(1982)}]{blandford1982}
{Blandford}, R.~D. \& {Payne}, D.~G. 1982, \mnras, 199, 883

\bibitem[{{Camenzind}(1990)}]{camenzind1990}
{Camenzind}, M. 1990, in Reviews in Modern Astronomy, Vol.~3, Reviews in Modern
  Astronomy, ed. {G.~Klare}, 234--265

\bibitem[{{Chandrasekhar} \& {Fermi}(1953)}]{chandrasekhar1953}
{Chandrasekhar}, S. \& {Fermi}, E. 1953, \apj, 118, 113

\bibitem[{{Cortes} {et~al.}(2008){Cortes}, {Crutcher}, {Shepherd}, \&
  {Bronfman}}]{cortes2008}
{Cortes}, P.~C., {Crutcher}, R.~M., {Shepherd}, D.~S., \& {Bronfman}, L. 2008,
  \apj, 676, 464

\bibitem[{{Cortes} {et~al.}(2005){Cortes}, {Crutcher}, \&
  {Watson}}]{cortes2005}
{Cortes}, P.~C., {Crutcher}, R.~M., \& {Watson}, W.~D. 2005, \apj, 628, 780

\bibitem[{{Crutcher}(1999)}]{crutcher1999}
{Crutcher}, R.~M. 1999, \apj, 520, 706

\bibitem[{{Crutcher}(2005)}]{crutcher2005}
{Crutcher}, R.~M. 2005, in IAU Symposium, ed. R.~{Cesaroni}, M.~{Felli},
  E.~{Churchwell}, \& M.~{Walmsley}, 98--107

\bibitem[{{Draine} {et~al.}(2007){Draine}, {Dale}, {Bendo}, {Gordon}, {Smith},
  {Armus}, {Engelbracht}, {Helou}, {Kennicutt}, {Li}, {Roussel}, {Walter},
  {Calzetti}, {Moustakas}, {Murphy}, {Rieke}, {Bot}, {Hollenbach}, {Sheth}, \&
  {Teplitz}}]{draine2007}
{Draine}, B.~T., {Dale}, D.~A., {Bendo}, G., {et~al.} 2007, \apj, 663, 866

\bibitem[{{Girart} {et~al.}(2009){Girart}, {Beltr{\'a}n}, {Zhang}, {Rao}, \&
  {Estalella}}]{girart2009}
{Girart}, J.~M., {Beltr{\'a}n}, M.~T., {Zhang}, Q., {Rao}, R., \& {Estalella},
  R. 2009, Science, 324, 1408

\bibitem[{{Girart} {et~al.}(1999){Girart}, {Crutcher}, \& {Rao}}]{girart1999}
{Girart}, J.~M., {Crutcher}, R.~M., \& {Rao}, R. 1999, \apjl, 525, L109

\bibitem[{{Girart} {et~al.}(2006){Girart}, {Rao}, \& {Marrone}}]{girart2006}
{Girart}, J.~M., {Rao}, R., \& {Marrone}, D.~P. 2006, Science, 313, 812

\bibitem[{{Goldreich} \& {Kylafis}(1981)}]{goldreich1981}
{Goldreich}, P. \& {Kylafis}, N.~D. 1981, \apjl, 243, L75

\bibitem[{{Goldreich} \& {Kylafis}(1982)}]{goldreich1982}
{Goldreich}, P. \& {Kylafis}, N.~D. 1982, \apj, 253, 606

\bibitem[{{Hildebrand}(1983)}]{hildebrand1983}
{Hildebrand}, R.~H. 1983, \qjras, 24, 267

\bibitem[{{Kylafis}(1983)}]{kylafis1983}
{Kylafis}, N.~D. 1983, \apj, 275, 135

\bibitem[{{Lai} {et~al.}(2003){Lai}, {Girart}, \& {Crutcher}}]{lai2003}
{Lai}, S., {Girart}, J.~M., \& {Crutcher}, R.~M. 2003, \apj, 598, 392

\bibitem[{{Marrone} {et~al.}(2006){Marrone}, {Moran}, {Zhao}, \&
  {Rao}}]{marrone2006}
{Marrone}, D.~P., {Moran}, J.~M., {Zhao}, J., \& {Rao}, R. 2006, \apj, 640, 308

\bibitem[{{Marrone} \& {Rao}(2008)}]{marrone2008}
{Marrone}, D.~P. \& {Rao}, R. 2008, in SPIE Conf.~Series, Vol. 7020

\bibitem[{{Mouschovias} \& {Paleologou}(1979)}]{mouschovias1979}
{Mouschovias}, T.~C. \& {Paleologou}, E.~V. 1979, \apj, 230, 204

\bibitem[{{Ostriker} {et~al.}(2001){Ostriker}, {Stone}, \&
  {Gammie}}]{ostriker2001}
{Ostriker}, E.~C., {Stone}, J.~M., \& {Gammie}, C.~F. 2001, \apj, 546, 980

\bibitem[{{Sault} {et~al.}(1995){Sault}, {Teuben}, \& {Wright}}]{sault1995}
{Sault}, R.~J., {Teuben}, P.~J., \& {Wright}, M.~C.~H. 1995, in ASP Conf. Ser.
  77, 433

\bibitem[{{Sridharan} {et~al.}(2002){Sridharan}, {Beuther}, {Schilke},
  {Menten}, \& {Wyrowski}}]{sridha}
{Sridharan}, T.~K., {Beuther}, H., {Schilke}, P., {Menten}, K.~M., \&
  {Wyrowski}, F. 2002, \apj, 566, 931

\bibitem[{{Stahler} \& {Palla}(2005)}]{stahler2005}
{Stahler}, S.~W. \& {Palla}, F. 2005, {The Formation of Stars} (ISBN
  3-527-40559-3.~Wiley-VCH)

\bibitem[{{Surcis} {et~al.}(2009){Surcis}, {Vlemmings}, {Dodson}, \& {van
  Langevelde}}]{surcis2009}
{Surcis}, G., {Vlemmings}, W.~H.~T., {Dodson}, R., \& {van Langevelde}, H.~J.
  2009, \aap, 506, 757

\bibitem[{{Tang} {et~al.}(2009){Tang}, {Ho}, {Koch}, {Girart}, {Lai}, \&
  {Rao}}]{tang2009a}
{Tang}, Y., {Ho}, P.~T.~P., {Koch}, P.~M., {et~al.} 2009, \apj, 700, 251

\bibitem[{{Troland} \& {Crutcher}(2008)}]{troland2008}
{Troland}, T.~H. \& {Crutcher}, R.~M. 2008, \apj, 680, 457

\bibitem[{{Vlemmings}(2008{\natexlab{a}})}]{vlemmings2008b}
{Vlemmings}, W.~H.~T. 2008{\natexlab{a}}, \aap, 484, 773

\bibitem[{{Vlemmings}(2008{\natexlab{b}})}]{vlemmings2008}
{Vlemmings}, W.~H.~T. 2008{\natexlab{b}}, in ASP Conference Series, Vol. 387,
  Massive Star Formation: Observations Confront Theory, ed. H.~{Beuther},
  H.~{Linz}, \& T.~{Henning}, 117

\bibitem[{{Vlemmings} {et~al.}(2010){Vlemmings}, {Surcis}, {Torstensson}, \&
  {van Langevelde}}]{vlemmings2010}
{Vlemmings}, W.~H.~T., {Surcis}, G., {Torstensson}, K.~J.~E., \& {van
  Langevelde}, H.~J. 2010, \mnras, 404, 134

\bibitem[{{Williams} {et~al.}(2004){Williams}, {Fuller}, \&
  {Sridharan}}]{williams2004}
{Williams}, S.~J., {Fuller}, G.~A., \& {Sridharan}, T.~K. 2004, \aap, 417, 115

\end{thebibliography}

\begin{table}[htb]
\caption{Additional online table: Polarization measurements with respect to 0/0 position R.A.\,(J2000.0) 18:11:51.40 Dec.\,(J2000.0) -17:31:28.5}
\begin{tabular}{lrr}
\hline \hline
Offset & $p$ & $\phi$ \\
($''$) & (\%)& (deg) \\
\hline
\multicolumn{3}{c}{Dust emission}\\
\hline
1.00,-2.25 & 5.6 & -48  \\
1.50,-1.75 & 4.2 & -66  \\
1.00,-1.75 & 3.7 & -68  \\ 
0.00,-1.75 & 2.5 & -47  \\
0.50,-1.25 & 1.4 & -42  \\
0.00,-1.25 & 1.6 & -43  \\
1.00,-0.75 & 1.0 & -35  \\
0.50,-0.75 & 1.0 & -38  \\
2.50,-0.25 & 5.7 &  33  \\
1.50,-0.25 & 1.4 & -24  \\
1.00,-0.25 & 1.1 & -36  \\
2.50,0.25  & 4.7 &  20  \\
2.00,0.25  & 1.6 & -12  \\
1.50,0.25  & 1.7 & -27  \\
1.00,0.25  & 1.7 & -45  \\
2.50,0.75  & 6.1 &  1  \\
\hline
\multicolumn{3}{c}{CO(3--2) at 40\,km\,s$^{-1}$}\\
\hline
  2.00,-12.25 &   8 &  -30 \\ 
  0.50,-12.25 &   8 &   37 \\ 
...\\
\hline \hline
\end{tabular}
\label{table}
\end{table}

\end{document}